\documentclass[aps,prd,notitlepage,showpacs,nofootinbib,superscriptaddress,12pt]{revtex4-1}

\usepackage[utf8]{inputenc} 

\usepackage{amsmath}
\usepackage{amssymb}
\usepackage{float}
\usepackage{comment}
\usepackage{slashed}
\usepackage[normalem]{ulem}
\usepackage{bbold}
\usepackage{cancel}
\usepackage{graphicx}
\usepackage{multirow}
\usepackage{rotating}
\usepackage{caption}
\usepackage{subcaption}
\usepackage{tabulary}

\newcolumntype{K}[1]{>{\centering\arraybackslash}m{#1}}
\def\gsim{\raise0.3ex\hbox{$\;>$\kern-0.75em\raise-1.1ex\hbox{$\sim\;$}}}
\def\lsim{\raise0.3ex\hbox{$\;<$\kern-0.75em\raise-1.1ex\hbox{$\sim\;$}}}

\usepackage[usenames,dvipsnames]{color}

\newcommand {\ignore}[1]{}

\usepackage[colorinlistoftodos]{todonotes}
\usepackage[colorlinks=true,linkcolor=darkred,citecolor=darkred,urlcolor=darkred, pdfborder={0 0 0}]{hyperref}
\usepackage[normalem]{ulem}

\usepackage{caption}
\usepackage{subcaption}
\definecolor{linkcolor}{rgb}{0,0,0.5}

\definecolor{darkgreen}{rgb}{0,0.5,0}
\definecolor{darkred}{rgb}{0.6,0,0}
\definecolor{brown}{rgb}{0.59, 0.29, 0.0}
\definecolor{mightnightblue}{RGB}{25,25,112}

\def\vev#1{\left\langle #1\right\rangle}
\def\SM{$\text{SU}(3)_c \otimes \text{SU}(2)_L \otimes \text{U}(1)_Y$ }

\def\znbb {$\rm 0\nu\beta\beta$ }

\newcommand{\AddrCFTP}{%
Departamento de F\'{\i}sica and CFTP, Instituto Superior T\'ecnico, Universidade de Lisboa, Av. Rovisco Pais 1, 1049-001 Lisboa, Portugal}

\newcommand{\AddrBhopal}{Department of Physics, Indian Institute of Science Education and Research - Bhopal, Bhopal Bypass Road, Bhauri, Bhopal 462066, India}
\newcommand{\AddrAHEP}{%
  AHEP Group, Institut de F\'{i}sica Corpuscular --
  CSIC/Universitat de Val\`{e}ncia, Parc Cient\'ific de Paterna.\\
 C/ Catedr\'atico Jos\'e Beltr\'an, 2 E-46980 Paterna (Valencia) - SPAIN}
\usepackage[force]{feynmp-auto}

\begin{document}

\title{\boldmath \color{BrickRed} Axion Paradigm with Color-Mediated Neutrino Masses}

\author{A. Batra}\email{aditya.batra@tecnico.ulisboa.pt}
\affiliation{\AddrCFTP}

\author{H.~B. C\^amara}\email{henrique.b.camara@tecnico.ulisboa.pt}
\affiliation{\AddrCFTP}

\author{ F.~R. Joaquim}\email{filipe.joaquim@tecnico.ulisboa.pt}
\affiliation{\AddrCFTP}

\author{R. Srivastava}\email{rahul@iiserb.ac.in}
\affiliation{\AddrBhopal}

\author{J.~W.~F. Valle}\email{valle@ific.uv.es}
\affiliation{\AddrAHEP}

\begin{abstract}
\vspace{0.5cm}
  We propose a generalized Kim-Shifman-Vainshtein-Zakharov-type axion framework in which colored fermions and scalars act as two-loop Majorana neutrino-mass mediators. The global Peccei-Quinn symmetry under which exotic fermions are charged solves the strong CP problem. Within our general proposal, various setups can be distinguished by probing the axion-to-photon coupling at helioscopes and haloscopes. We also comment on axion dark-matter production in the early Universe.
\end{abstract}

\maketitle
\noindent

\section{Introduction}
%
Despite its remarkable success, the Standard Model~(SM) fails to account for neutrino masses~\cite{Kajita:2016cak,McDonald:2016ixn} and cosmological dark matter~(DM)~\cite{Bertone:2004pz,Planck:2018vyg}, 
new physics being required to explain both phenomena. Most SM extensions either focus on only one of these problems or, when both are addressed, their solutions are not directly connected.
However, there have been many attempts to relate DM to neutrino mass generation. For example, in scotogenic scenarios neutrino masses arise radiatively from the exchange of DM states~\cite{Tao:1996vb,Ma:2006km}, a paradigm that has been further developed in many recent studies -- see e.g. Refs.~\cite{Hirsch:2013ola,Toma:2013zsa,Vicente:2014wga,Ahriche:2016cio,Reig:2018mdk,Reig:2018ztc,Barreiros:2022aqu,Chun:2023vbh,Diaz:2016udz,Merle:2016scw,Avila:2019hhv,Restrepo:2019ilz,Karan:2023adm}. An interesting alternative is to consider dark sectors which trigger lepton number violation and neutrino mass generation within low-scale seesaw schemes~\cite{Mandal:2019oth,Batra:2023bqj,CarcamoHernandez:2023atk}.

Another long-standing drawback of the SM concerns the strong charge-parity~(CP) symmetry violation in Quantum Chromodynamics~(QCD), encoded by the (arbitrary) phase parameter $\bar{\theta}$. The fact that severe experimental constraints on the neutron electric dipole moment~\cite{PhysRevD.92.092003,PhysRevLett.97.131801} force $\left|\overline{\theta}\right| < 10^{-10}$ is known as the strong CP problem. One of the most elegant solutions to this problem is the Peccei-Quinn~(PQ) mechanism~\cite{PhysRevLett.38.1440,PhysRevD.16.1791} 
based on a classical QCD-anomalous $\text{U}(1)_{\text{PQ}}$ symmetry. Spontaneous $\text{U}(1)_{\text{PQ}}$ breaking leads to a pseudo Goldstone boson -- the axion $a$~\cite{PhysRevLett.40.223,PhysRevLett.40.279}.
The ground state of $a$ turns out to be such that it effectively sets $\overline{\theta} = 0$, providing a dynamical solution to the strong CP problem. Among the plethora of axion models, two main sets can be distinguished: the Dine-Fischler-Srednicki-Zhitnitsky~(DFSZ)~\cite{Zhitnitsky:1980tq,Dine:1981rt} and the Kim-Shifman-Vainshtein-Zakharov~(KSVZ)~\cite{Kim:1979if,Shifman:1979if} invisible-axion scenarios. In the former, SM quarks are charged under $\text{U}(1)_{\text{PQ}}$, while in the latter the PQ-charged fields are exotic quarks~(for a review see Ref.~\cite{DiLuzio:2020wdo}).

Axions, which may be produced non-thermally in the early Universe via the so-called misalignment mechanism~\cite{Preskill:1982cy,Abbott:1982af,Dine:1982ah}, can also be excellent alternatives~\cite{DiLuzio:2020wdo} to weakly interacting massive particle DM~\cite{Arcadi:2017kky}.
It is then tempting to embed the axion paradigm in frameworks that simultaneously provide an explanation for small neutrino masses.
This has been explored in the literature recently, for example by realizing the DFSZ or KSVZ axion within the type-I seesaw.
  Technically, natural setups that also address other SM shortcomings such as the baryon asymmetry of the Universe and inflation were proposed~\cite{Salvio:2015cja,Ballesteros:2016euj,Ballesteros:2016xej,Clarke:2015bea,Sopov:2022bog}.
  In this Letter, we suggest a new idea in which neutrino masses are generated at the quantum level via colored mediators that also provide a solution to the strong CP problem. This new class of KSVZ-type axion models connects three otherwise unrelated issues:  small neutrino masses, the strong CP problem and DM.

\begin{table}[t!]
\setlength{\tabcolsep}{-1pt}
\renewcommand*{\arraystretch}{1.6}
	\centering
	\begin{tabular}{| K{1.5cm} | K{5cm} | K{1.5cm} | K{2.5cm} |}
		\hline
 Fields&\SM&    U($1$)$_{\text{PQ}}$ & Multiplicity\\
		\hline \hline
$\Psi_L$&($(p,q),2 n \pm 1, 0$)& $\omega$  & $n_\Psi$\\
$\Psi_R$&($(p,q),2 n \pm 1, 0$)&  $0$ & $n_\Psi$ \\
		\hline \hline
$\sigma$&($\mathbf{1},\mathbf{1}, 0$)& $\omega$ & 1\\
$\eta$&($(p,q),2 n, 1/2$)& $0$ & $n_\eta$\\
$\chi$&($(p,q),2 n \pm 1, 0$)& $0$ & $n_\chi$ \\
\hline
	\end{tabular}
	\caption{
          Matter content and quantum numbers for our new KSVZ models with two-loop neutrino masses.
          Here, $\omega$ is the PQ charge, and $n = 1,2, \cdots$, $p>q = 0,1,2, \cdots$.}
	\label{tab:general} 
\end{table}
%
\section{Framework}
%
The original KSVZ model~\cite{Kim:1979if,Shifman:1979if} extends the SM with vectorlike fermions $\Psi_{L,R}$ in the fundamental representation of SU(3)$_c$, singlets under SU(2)$_L$, and with $Y=0$. A complex scalar singlet $\sigma$ breaks a U($1$)$_{\text{PQ}}$ symmetry spontaneously, providing mass to those exotic fermions. The phase of $\sigma$ corresponds to the axion field $a$. The fact that left-handed and right-handed exotic fermions carry different PQ charges ensures the anomalous axion-gluon coupling, required to solve the strong CP problem. 

In this Letter, we show that generic $\Psi_{L,R}$ fields in the SU($3)_c$ complex representation $(p,q)$ with $p>q = 0,1,2, \cdots$ can act as neutrino-mass mediators at the two-loop level. Two scalars $\eta,\chi$ with those same SU($3)_c$ transformation properties are also required to put our mechanism at work. Both $\Psi_{L,R}$ and $\chi$ are hyperchargeless and transform as the same odd SU($2)_L$ representation denoted by $2n \pm 1$. In contrast, $\eta$ has $Y=1/2$ and transforms as an even SU($2)_L$ representation denoted by 
$2n$. As in the original KSVZ prescription, the complex scalar singlet $\sigma$ with nonzero PQ charge $\omega$ is responsible for U($1$)$_{\text{PQ}}$ breaking, giving rise to the axion and to $\Psi_{L,R}$ masses (note that only $\Psi_L$ carries PQ charge $\omega$,
$\Psi_R$ is neutral). Table \ref{tab:general} lists all the new fields and their transformation properties under the SM and PQ symmetries.

The relevant new Yukawa terms are given by
\begin{align}
- \mathcal{L}_{\text{Yuk.}} &\supset \mathbf{Y}_{\Psi} \overline{\Psi_L} \Psi_R \sigma
                           + \frac{1}{2} \mathbf{Y}_{\chi_j} {\Psi^T_R}~C~\chi_j \Psi_R + \mathbf{Y}_i \overline{L} \; \eta^\ast_i \Psi_R + \text{H.c.} \; ,
\label{eq:LYukgen}
\end{align}
where $L$ denotes the SM lepton doublet. For simplicity, from now on we omit color and SU(2)$_L$ indices. The multiplicities of $\Psi$, $\eta$ and $\chi$ are $n_\Psi$, $n_\chi$ and $n_\eta$, respectively. Thus $\mathbf{Y}_{\Psi}$, $\mathbf{Y}_{\chi_j}$ and $\mathbf{Y}_i$ are $n_\Psi \times n_\Psi$, $n_\Psi \times n_\Psi$ and $3\times n_\Psi$ complex Yukawa matrices, respectively,
with $i=1, \cdots, n_\eta$ and $j=1, \cdots, n_\chi$.

The key scalar-potential terms responsible for neutrino mass generation are
\begin{align}
    V & \supset \mu_{ijk} \chi_i\chi_j\chi_k + \kappa_{ij} \eta_i^\dagger \Phi  \chi_j + \lambda_{ijk} \Phi^\dagger \eta_i \chi_j \chi_k + \text{H.c.} \; ,
    \label{eq:Vneutrino}
\end{align}
where $\Phi=(\phi^+ , \phi^0)^T$ is the SM Higgs doublet. To preserve the SU($3)_c$ symmetry the colored scalars $\eta$ and $\chi$ must not acquire a vacuum expectation value, so that the only vacuum expectation values are $\vev{\sigma} = v_\sigma/\sqrt{2}$ breaking U($1$)$_{\text{PQ}}$, and $\vev\phi^0 = v/\sqrt{2}\simeq 174$~GeV triggering electroweak symmetry breaking. 

\section{Strong CP problem}
%
The PQ field $\sigma = (v_\sigma + \rho) \exp(i a /v_\sigma)/\sqrt{2}$ contains the axion $a$ and the radial mode $\rho$. Once~$\sigma$ develops a non-zero $v_\sigma$, the PQ symmetry is spontaneously broken at a scale $f_{\text{PQ}} = \langle \sigma \rangle= v_\sigma/\sqrt{2}$, leading to the axion decay constant
\begin{equation}
    f_a = \frac{f_{\text{PQ}}}{N} = \frac{v_{\sigma}}{\sqrt{2} N} \; ,
    \label{eq:axionfa}
\end{equation}
where $N$ is the color anomaly factor.  The up-to-date QCD axion mass at next-to-leading order is~\cite{GrillidiCortona:2015jxo}
\begin{equation}
    m_a = 5.70(7) \left(\frac{10^{12} \text{GeV}}{f_a}\right) \mu \text{eV} \; .
    \label{eq:axionmass}
\end{equation}
This relation between $m_a$ and $f_a$ is a model-independent prediction of the QCD axion if the only explicit breaking of the PQ symmetry is by nonperturbative QCD effects. To be viable, the axion solution to the strong CP problem requires a nonvanishing anomaly factor $N$ to ensure an axion-gluon coupling. For the models in Table~\ref{tab:general} we get 
\begin{align}
N &= 2 \, n_\Psi \, \omega \, (2 n \pm 1) \; T(p,q) \; ,
\label{eq:Nmodel}
\end{align}
with $T(p,q)$ the Dynkin index of the SU($3)_c$ representation $(p,q)$. As expected, $N$ depends on the multiplicity of the colored fermions $n_\Psi$ and on the $\Psi_L$ PQ charge $\omega$.

    \begin{figure*}[t!]
        \centering
        \includegraphics[scale=0.58]{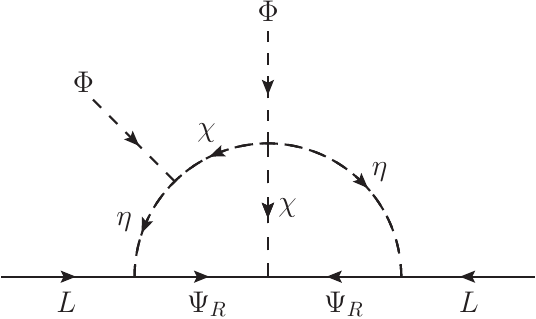} \hspace{+0.1cm}
        \includegraphics[scale=0.58]{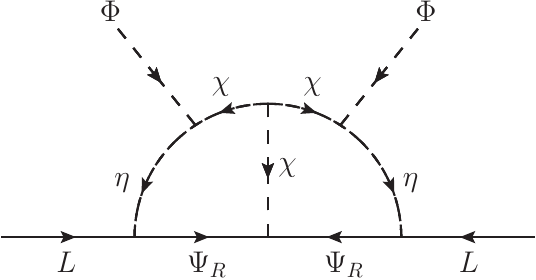} \hspace{+0.1cm}
        \includegraphics[scale=0.58]{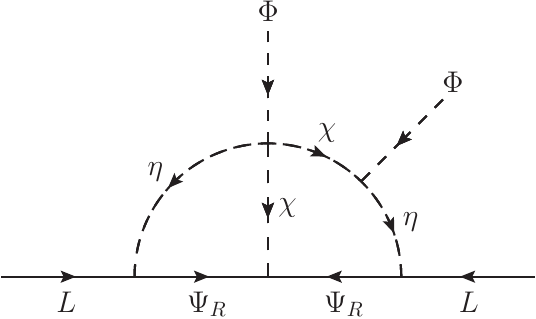}
        \caption{Two-loop diagrams for neutrino-mass generation mediated by the colored particles of Table~\ref{tab:general}.}
    \label{fig:neutrino}
    \end{figure*}
%
\section{Neutrino mass generation}
%
With the Yukawa and scalar interactions of Eqs.~\eqref{eq:LYukgen} and~\eqref{eq:Vneutrino}, two-loop Majorana neutrino masses arise from the diagrams in Fig.~\ref{fig:neutrino}
\footnote{
  Two-loop neutrino mass diagrams have been systematically classified in Refs.~\cite{AristizabalSierra:2014wal,Cai:2017jrq}.  Our topologies also arise in many specific schemes, such as those in Refs.~\cite{Cheng:1980qt,Zee:1985id,Babu:1988ki,Babu:2002uu,Ma:2007gq,Bonilla:2016diq,Aoki:2014cja,Okada:2014qsa,Ho:2016aye,Ho:2017fte,Baek:2017qos}.}.

Neutrino masses are mediated by colored particles, $\Psi$, $\eta$ and $\chi$, transforming under the same SU($3)_c$ representation. Moreover, since $L$ and $\Phi$ are SU($2)_L$ doublets, $\eta$ must lie in an even SU($2)_L$ representation, whereas $\Psi$ and $\chi$ must be in an odd representation. The coupling between $L$ and $\Psi$ requires $\eta$ to have $Y=1/2$. In our scenario $\Psi$ and $\chi$ carry no hypercharge, ensuring the Majorana nature of light neutrinos. 

Among all generic scenarios in Table~\ref{tab:general}, the simplest consistent realization of our idea is for $\Psi_{L,R}$ and $\eta,\chi$ to transform as triplets of SU(3)$_{c}$.
Since the SU(3) invariant coming from $(p,q) \otimes (p,q) \otimes (p,q)$ for $(p,q) \equiv \mathbf{3}$ is antisymmetric, the minimal required multiplicity is $n_\Psi = n_\chi = 2$,~$n_\eta=1$.
For symmetric contractions, $n_\chi$ can also be as small as 1. Concerning SU(2)$_{L}$, $\eta$ is a doublet while $\Psi$ and $\chi$ are singlets. 
For $\omega=1/2$, this setup predicts $N=1$, just as in the original KSVZ model~\cite{Kim:1979if,Shifman:1979if}. This minimal scenario simply extends the original KSVZ proposal with extra colored scalars $\eta$ and $\chi$, which mediate neutrino-mass generation. The resulting light-neutrino mass matrix is
\begin{align}
    (m_\nu)_{\alpha \beta} &= \frac{N_\text{c}}{(16 \pi^2)^2} \ \tilde{Y}_{a\alpha}^j \ (\tilde{Y}_{\chi})_{a b}^k \ \tilde{Y}_{b \beta}^l \ \tilde{\mu}_{j k l} \ \mathcal{I}_{a b}^{j k l} \; ,
     \label{eq:mnu}
    \end{align}
where $j,k,l=1, \cdots, 6$ and $a,b=1,2$. $N_c=6$ is the color factor, $\tilde{Y}$ and $\tilde{Y}_\chi$ are Yukawa couplings, while $\tilde{\mu}$ denotes the cubic scalar couplings of Eq.~\eqref{eq:Vneutrino}, all written now in the mass basis. The loop function $\mathcal{I}_{a b}^{j k l}$ can be found in Refs.~\cite{AristizabalSierra:2014wal,Aoki:2014cja}. The above result can be estimated by
 \begin{align}
    (m_\nu)_{\alpha \beta} & \sim \ 0.1 \ \text{eV} \left(\frac{\tilde{Y}_{a\alpha}^j \ (\tilde{Y}_{\chi})_{a b}^k \ \tilde{Y}_{b \beta}^l}{10^{-3}} \right) \ \left(\frac{\tilde{\mu}_{j k l}}{10^8 \ \text{GeV}}\right) \left(\frac{v}{246 \ \text{GeV}}\right)^2 \ \left(\frac{10^8 \ \text{GeV}}{m_\zeta}\right)^2 \; ,
      \label{eq:mnuapprox}
\end{align}
where $m_\zeta = \sqrt{\lambda_{\text{eff}}} f_{\text{PQ}}$ is an effective colored scalar mass scale running in the loop with $\lambda_{\text{eff}}$ being some quartic coupling parameter. A typical value for the PQ breaking scale is $f_{\text{PQ}} \sim 10^{12}$ GeV, so that axions account for the observed DM relic abundance. Hence, the scalars are expected to be heavy. The smallness of $\tilde{Y}_\chi$ and $\tilde{\mu}_{j k l}$ in Eq.~(\ref{eq:mnu}) is symmetry-protected in
t'Hooft's sense~\cite{tHooft:1979rat}, as the Lagrangian acquires an additional U(1) symmetry in their absence. Note also that, with only two copies of $\Psi$ ($n_\Psi=2$), one of the three light neutrinos is predicted to be massless due to the missing partner nature~\cite{Schechter:1980gr} of the underlying radiative seesaw mechanism. Charged lepton flavor violating processes would be mediated at one-loop by the charged colored scalars and exotic fermions, but with very small rates~\footnote{Up to color factors, expressions for such rates resemble those of similar scenarios, e.g. scotogenic models~\cite{Toma:2013zsa,Vicente:2014wga,Ahriche:2016cio,Reig:2018mdk,Mandal:2019oth,Barreiros:2022aqu,Chun:2023vbh}.}. 

In its minimal version, the above scenario implies that there is no cancellation in the \znbb amplitude, even for normally-ordered neutrino masses~\cite{Reig:2018ztc,Barreiros:2018bju,Avila:2019hhv}. The resulting regions allowed by oscillation data~correlate with the only free parameter available, i.e. the relative neutrino Majorana phase. One finds that, for inverted ordering, rates fall inside the expected sensitivities of the next round of experiments~\cite{KamLAND-Zen:2022tow,GERDA:2019ivs,GERDA:2020xhi,Adams:2022jwx}.
These would not only prove the Majorana nature by the black-box theorem~\cite{Schechter:1981bd}, but could also ultimately determine the Majorana phase~\cite{Branco:2002ie}.

\begin{table}[t!]
\setlength{\tabcolsep}{-1pt}
\renewcommand*{\arraystretch}{1.5}
	\centering
\begin{tabular}{| K{1.5cm} | K{1.5cm} | K{1.5cm} | K{1.5cm} |  K{1.5cm} | K{1.5cm} |  K{1.5cm} |}
        \hline
\multicolumn{2}{|c|}{\multirow{2}{*}{$E/N$}} & \multicolumn{5}{c|}{SU($2)_L$}\\
		\cline{3-7}
\multicolumn{2}{|c|}{} & $\mathbf{3}$ & $\mathbf{5}$ & $\mathbf{7}$ & $\mathbf{9}$ & $\mathbf{11}$ \\
		\hline 
        \multirow{5}{*}{\rotatebox[origin=c]{90}{SU($3)_c$}}
&$\mathbf{3}$ & 4 & 12 & 24 & 40 & 60 \\
&$\mathbf{6}$ & 8/5 & 24/5 & 48/5 & 16 & 24 \\
&$\mathbf{10}$ & 8/9 & 8/3 & 16/3 & 80/9 & 40/3 \\
&$\mathbf{15}$ & 1 & 3 & 6 & 10 & 15 \\
&$\mathbf{15'}$ & 4/7 & 12/7 & 24/7 & 40/7 & 60/7 \\
        \hline
\end{tabular}
\caption{
  $E/N$ values for various $\mathrm{SU(3)}_c \otimes \mathrm{SU(2)}_L$ representation choices for $\Psi$ [see Table~\ref{tab:general} and Eq.~\eqref{eq:E}].}
\label{tab:EN}
\end{table}
%
\section{Probing the axion-to-photon coupling}
%
Indirect astrophysical and cosmological observations, as well as laboratory searches~(for reviews see Refs.~\cite{DiLuzio:2020wdo,Adams:2022pbo}), constrain the axion parameter space due to its couplings to photons, nucleons and electrons.
We now examine how to probe the various scenarios of Table~\ref{tab:general} through their corresponding axion-to-photon coupling $g_{a \gamma \gamma}$.

In the KSVZ setup, the only chiral fermions charged under U($1$)$_{\text{PQ}}$ are the new exotic fermions. Therefore, there are no model-dependent contributions to the axion coupling to nucleons and electrons. Using next-to-leading-order chiral Lagrangian techniques, one obtains~\cite{GrillidiCortona:2015jxo} 
\begin{align}
    g_{a \gamma \gamma} &= \frac{\alpha_e}{2 \pi f_a} \left[\frac{E}{N} - 1.92(4) \right] \; ,
    \label{eq:gagg}
\end{align}
where $E$ and $N$ are the model-dependent electromagnetic and color anomaly factors, respectively. For our class of models in Table~\ref{tab:general}, we have
\begin{equation}
    \frac{E}{N} = \frac{d(p,q)}{(2 n \pm 1) T(p,q)} \sum_{j=0}^{2n \pm 1 - 1} \left(\frac{2n \pm 1 - 1}{2} - j \right)^2 \; ,
    \label{eq:E}
\end{equation}
with $d(q,p)$ being the dimension of SU(3)$_c$ representation. One sees that $E/N = 0$, as long as the hyperchargeless $\Psi_{L,R}$ are SU($2)_L$ singlets. For higher weak multiplet representations $E/N \neq 0$ -- see Table~\ref{tab:EN}.
\begin{figure}[!t]
    \centering
      \includegraphics[scale=0.5]{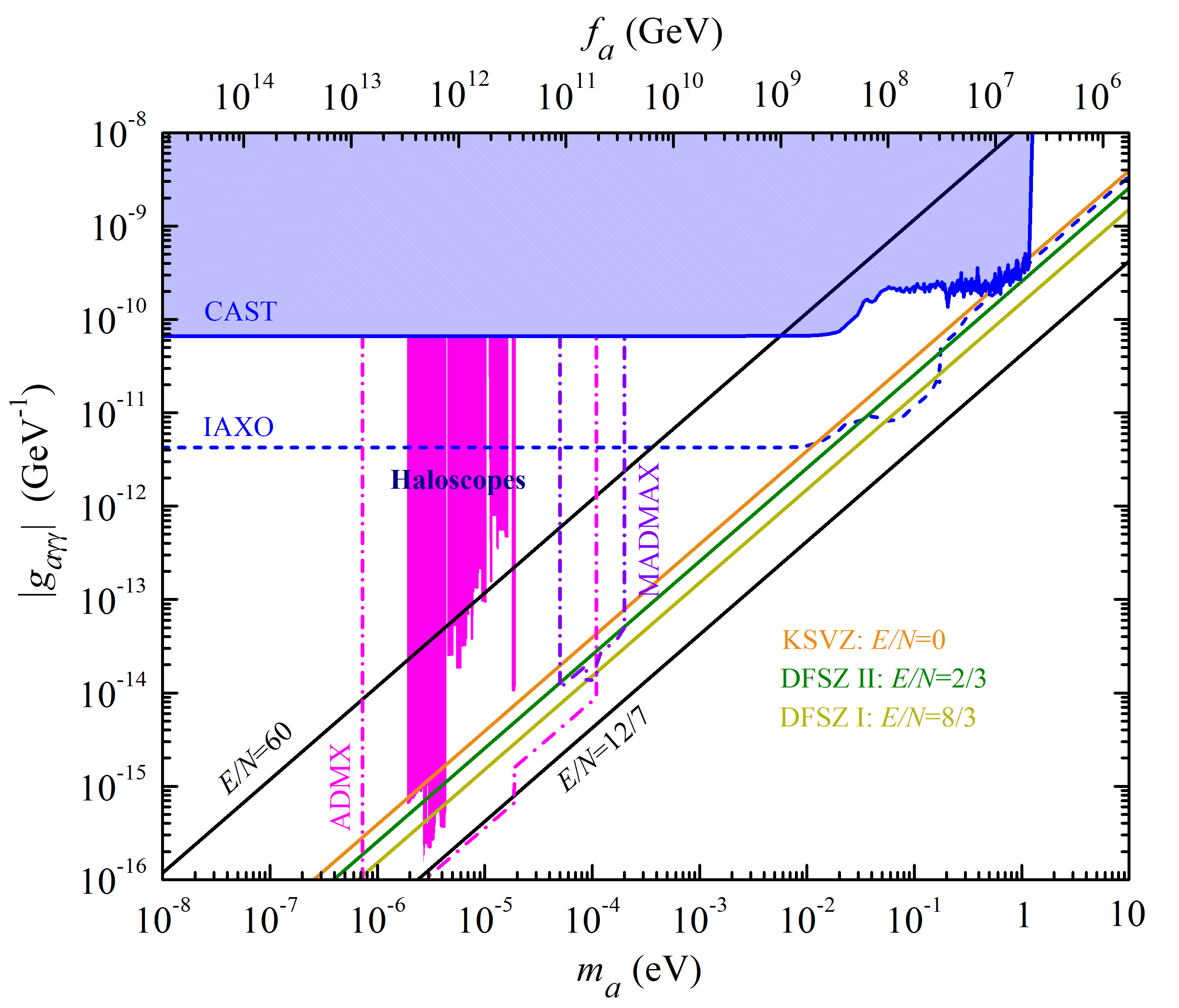}
      \caption{
        $|g_{a \gamma \gamma}|$ versus $m_a$ (bottom axis) and $f_a$ (top axis) [see Eqs.~\eqref{eq:axionmass} and~\eqref{eq:gagg}].
        The black lines correspond to $E/N$ values leading to maximum and minimum $|g_{a \gamma \gamma}|$ for the representations shown in Table~\ref{tab:EN}. 
        The KSVZ and DFSZ I and II predictions are indicated by the orange, light green, and dark green lines, respectively. Shaded regions are presently excluded, while dash-dotted lines delimit projected sensitivities of several helioscope and haloscope experiments -- see text for details.}
    \label{fig:gagg}
\end{figure}

In Fig.~\ref{fig:gagg}, we display by oblique solid lines the axion-photon coupling $|g_{a\gamma \gamma}|$ in terms of~$m_a$ (bottom axis) and~$f_a$ (top axis). The black lines delimit the band of $E/N$ values leading to the maximum and minimum $|g_{a\gamma \gamma}|$, corresponding to $E/N=60$ for~$\Psi \sim (\mathbf{3},\mathbf{11},0)$ and $E/N=12/7$ for~$\Psi \sim (\mathbf{15}^\prime,\mathbf{5},0)$, respectively [see Eq.~\eqref{eq:gagg} and Table~\ref{tab:EN}]. The $|g_{a\gamma \gamma}|$ corresponding to the popular KSVZ and DFSZ-I and II schemes are shown by the solid orange, light green, and dark green lines, respectively. The minimal KSVZ model featuring two-loop neutrino masses predicts $E/N=0$ (solid orange line). 

In the same plot we show the current bounds and future sensitivities from helioscopes and haloscopes. The CAST helioscope experiment~\cite{CAST:2017uph} excludes the blue-shaded region, while haloscopes ADMX~\cite{ADMX:2018gho,ADMX:2019uok,ADMX:2021nhd}, RBF~\cite{DePanfilis:1987dk}, CAPP~\cite{CAPP:2020utb}, and HAYSTAC~\cite{HAYSTAC:2020kwv} exclude the magenta region. Projected sensitivities of IAXO~\cite{Shilon_2013}, ADMX~\cite{Stern:2016bbw}, and MADMAX~\cite{Beurthey:2020yuq} are indicated by the dashed blue, magenta, and purple contours, respectively. One sees that the future (2025) IAXO experiment is expected to probe $g_{a \gamma \gamma}$ down to $(10^{-12}-10^{-11}) \ \text{GeV}^{-1}$ reaching the popular QCD axion model predictions for $m_a \sim 0.1$ eV (blue-dashed contour).  Out of all haloscope experiments, the most impressive is ADMX, which has already reached the KSVZ and DFSZ QCD axion lines for masses $m_a \sim 3 \ \mu$eV. Upcoming ADMX (dash-dotted magenta contour) should probe the full landscape of QCD axion models for masses $1 \ \mu \text{eV} \lesssim m_a \lesssim 100 \ \mu$eV. Moreover, MADMAX (2024)~\cite{Beurthey:2020yuq} is projected to cover the region $50 \ \mu \text{eV} \lesssim m_a \lesssim 120 \ \mu$eV (dash-dotted purple contour).

\begin{figure}[!t]
    \centering
      \includegraphics[scale=0.5]{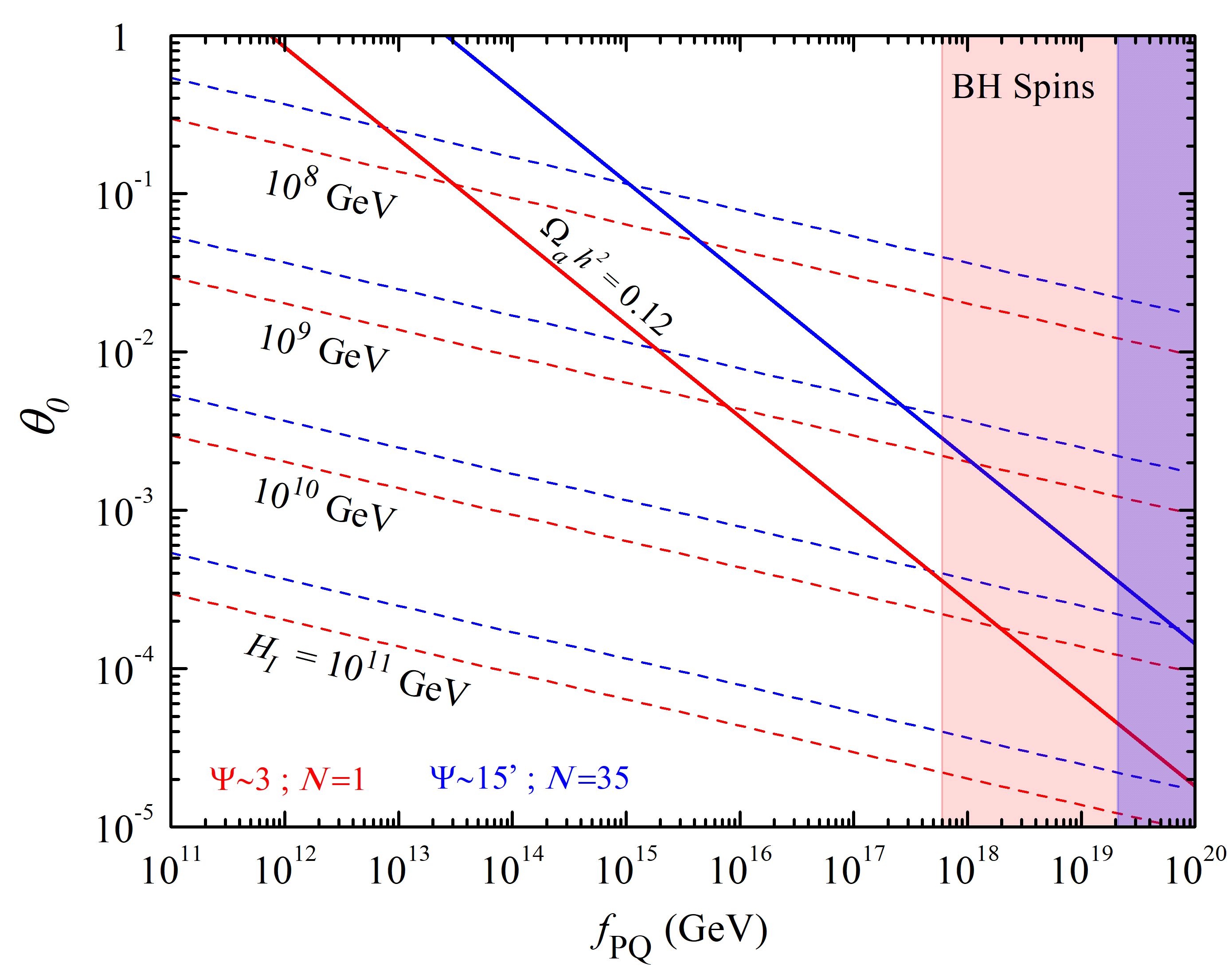}
      \caption{
        Misalignment angle $\theta_0$ as a function of $f_{\text{PQ}}$.
        In red (blue) we show the scenario for $\Psi \sim 3$ ($\Psi \sim 15^\prime$) under SU($3)_c$, singlet under SU($2)_L$, with $\omega=1/2$ and $n_\Psi=2$.
        Along the solid lines $\Omega_a h^2=0.12$, above (below) them we have DM over- or underabundance.
        Above the dashed lines the value of the inflationary scale $H_I$ lies below the indicated value [see Eq.~\eqref{eq:Iso}].
        Vertical bands are excluded by black hole superradiance.
      }
\label{fig:DM}
\end{figure}
%
\section{Axion dark matter and cosmology}
%
Axions are naturally light, weakly coupled with ordinary matter, cosmologically stable, and can be nonthermally produced in the early Universe. Indeed, they turn out to be an excellent DM candidate. In a preinflationary scenario, the PQ symmetry is broken before (or during) inflation and never restored during the reheating period of the Universe. Axion DM production occurs through the misalignment mechanism~\cite{Preskill:1982cy,Abbott:1982af,Dine:1982ah}, with axion relic abundance given by~\cite{DiLuzio:2020wdo}
\begin{equation}
   \Omega_a h^2 \simeq  \Omega_\text{CDM} h^2 \frac{\theta_0^2}{2.15^2} \left(\frac{f_a}{2 \times 10^{11} \ \text{GeV}} \right)^{\frac{7}{6}} \; ,
   \label{eq:relica}
\end{equation}
where the free parameter $|\theta_0| \in [0 , \pi)$ is the initial misalignment angle and the observed cold DM~(CDM) relic abundance obtained by Planck is
$\Omega_{\text{CDM}} h^2 = 0.1200 \pm 0.0012$~\cite{Planck:2018vyg}. For $\theta_0 \sim \mathcal{O}(1)$ and $f_a \sim 5 \times 10^{11} \ \text{GeV}$, axions can account for the full DM abundance. 

In a postinflationary scenario, where the PQ symmetry is broken after inflation, the observable Universe will be divided in patches with different values of the axion field (or $\theta$ phase). The initial misalignment angle is obtained through statistical average as $\left<\theta_0^2\right> \simeq 2.15^2$~\cite{DiLuzio:2020wdo}. Thus, if the axion makes up the full CDM, $\Omega_a h^2 = \Omega_{\text{CDM}} h^2$, so that $f_a$ is predicted as in Eq.~\eqref{eq:relica}. Hence, if only the misalignment mechanism is at play, $f_a \lesssim 2 \times 10^{11}$ GeV ensures that DM is not overproduced. However, the picture gets more complicated since topological defects (strings and domain walls) can also contribute to $\Omega_a h^2$~\cite{Bennett:1987vf,Levkov:2018kau,Gorghetto:2018myk,Buschmann:2019icd}~\footnote{Notice that $N=1$ avoids cosmological domain walls. This is achieved, for the models in Table~\ref{tab:general}, if the PQ charge is fixed to $\omega^{-1} = 2 n_\Psi \; (2 n \pm 1) \; T(p,q) $ [see Eq.~\eqref{eq:Nmodel}].}. Note that the lightest state stemming from the colored fields $\Psi$, $\eta$, or $\chi$ can be cosmologically stable, and it can be thermally produced after inflation~\footnote{After symmetry breaking, due to color and electromagnetic symmetries, the Lagrangian exhibits an unbroken accidental $\mathcal{Z}_3$ symmetry where $(\Psi,\eta,\chi) \rightarrow e^{i 2\pi/3} (\Psi,\eta,\chi)$, stabilizing the lightest colored state [see Eqs.~\eqref{eq:LYukgen} and~\eqref{eq:Vneutrino}].}. Searches in terrestrial, lunar, and meteoritic materials yield strong limits~\cite{Perl:2001xi,Perl:2009zz,Burdin:2014xma,Hertzberg:2016jie,Mack:2007xj}, practically ruling out such stable charged baryonic relics, unless some mechanism effectively suppresses their density or allows them to decay to ordinary matter~\footnote{Exotic fermions $\Psi$ with $Y \neq 0$ can possibly mix with ordinary quarks allowing scenarios free of stable colored or charged relics, leading to viable postinflationary axion DM. These cases are identified in the Supplemental Material. }~\cite{Nardi:1990ku,DiLuzio:2016sbl,DiLuzio:2017pfr}~\footnote{If electrically neutral, these relics might form viable bound-state DM~\cite{Reig:2018mdk,Reig:2018ztc}.}.

\vspace{+0.1cm}

Turning to the preinflationary scenario, we assume that the exotic fermion and scalar masses lie above the reheating temperature of the Universe, i.e. $m_{\Psi,\eta,\chi} > T_\text{RH}$. This is a reasonable assumption since their masses are proportional to $f_{\text{PQ}} \gg v$, and $T_{\text{RH}}$ is only bounded from below by big bang nucleosynthesis~\cite{deSalas:2015glj}, $T_{\text{RH}} \gsim 4.7 \ \text{MeV}$. This way, the abundance of stable baryonic or charged relics will be washed out during inflation, as well as topological defects. In preinflationary scenarios, the axion leaves an imprint in primordial fluctuations, reflected in the cosmic microwave background anisotropies and large-scale structure. The resulting isocurvature fluctuations are constrained by cosmic microwave background data~\cite{Beltran:2006sq}, leading to an upper bound on the inflationary scale $H_I$~\cite{DiLuzio:2016sbl}:
\begin{equation}
   H_I \lsim  \frac{0.9\times 10^7}{\Omega_a h^2/\Omega_\text{CDM} h^2} \left(\frac{\theta_0}{\pi} \frac{f_a}{ 10^{11} \ \text{GeV}} \right) \; \text{GeV} \; .
   \label{eq:Iso}
\end{equation}

In Fig.~\ref{fig:DM}, we display $\theta_0$ as a function of $f_{\text{PQ}}$ [see Eq.~\eqref{eq:axionfa}]. We highlight two cases, in red and blue, where the fermions, singlets under SU($2)_L$, transform as $\Psi \sim 3$ and $\Psi \sim 15^\prime$ under SU($3)_c$, respectively. We take $\omega=1/2$ and $n_\Psi=2$. Along the solid lines we have $\Omega_a h^2=0.12$ [see Eq.~\eqref{eq:relica}]. The region above these lines is excluded since it implies DM overabundance. Black hole superradiance sets $f_a \leq  6 \times 10^{17} \ \text{GeV}$~\cite{Arvanitaki:2014wva,Dafni:2018tvj} (shaded bands). Taking $\theta_0 \sim \mathcal{O}(1)$ leads to $f_a \gsim 5 \times 10^{11}$ GeV, a region currently being probed by haloscope experiments -- see Fig.~\ref{fig:gagg}. The dashed lines indicate different values of the inflationary scale $H_I$. Above these lines $H_I$ is below the indicated value, in agreement with the isocurvature bound of Eq.~\eqref{eq:Iso}. The allowed region for a given $H_I$ lies above the dashed and below the solid contours. Taking $\theta_0 \sim \mathcal{O}(1)$ and $\Omega_a h^2=0.12$, we get a low scale for inflation $H_I \lsim 10^7$ GeV (Planck currently probes $H_I \lsim 10^{13}$ GeV~\cite{Planck:2018vyg}). 

\section{Final remarks}
%
In this Letter, we proposed a connection between two seemingly unrelated facts: small neutrino masses and the strong CP problem. This was achieved within a novel class of KSVZ axion schemes, containing exotic colored fermions and scalars which act as neutrino-mass mediators at the two-loop level. The simplest realization of our proposal leads to promising \znbb decay predictions.

Different representation assignments of the new fields under the SM and PQ symmetries yield distinct axion-to-photon couplings. This provides a way to differentiate the various realizations of our scheme at future helioscope and haloscope experiments such as IAXO, ADMX, and MADMAX. 

Because of potentially dangerous colored relics, we have considered axion DM in the preinflationary scenario, where the PQ symmetry is broken before inflation. For an initial misalignment angle $\theta_0 \sim \mathcal{O}(1)$, axions can account for the full CDM budget, provided $f_a \sim5 \times 10^{11}$ GeV, a region currently under scrutiny at haloscopes. 

\begin{acknowledgments}
This research is supported by Funda\c{c}\~ao para a Ci\^encia e a Tecnologia (FCT, Portugal) through the projects CFTP-FCT Unit UIDB/00777/2020 and UIDP/00777/2020, CERN/FIS-PAR/0019/2021,
which are partially funded through POCTI (FEDER), COMPETE, QREN and EU.
J.W.F.V. is supported by Spanish grants PID2020-113775GB-I00 (AEI/10.13039/501100011033) and Prometeo CIPROM/2021/054 (Generalitat Valenciana). 
The work of A.B. and H.B.C. is supported by the PhD FCT grants UI/BD/154391/2023 and 2021.06340.BD, respectively. F.R.J. thanks the CERN Theoretical Physics Department for hospitality and financial support during the final stage of this work.
\end{acknowledgments}

\vspace{+0.5cm}

	\appendix
	
	\begin{center}
		\bf \large Supplemental Material
	\end{center}

\renewcommand\theequation{S\arabic{equation}}
\renewcommand\thetable{S\Roman{table}}
\renewcommand\thefigure{S\arabic{figure}}

\setcounter{equation}{0}
\setcounter{table}{0}
\setcounter{figure}{0}

	Here we present the possibility for the exotic fermions $\Psi$ having non-zero hypercharge, so that they can mix with SM quarks, and discuss its consequences. 
	The potential advantage of such a scenario is that the mixing with quarks can open decay channels for $\Psi$ suppressing the relic density in the current universe.
        In order for $\Psi$ to mix with SM quarks, these must transform as triplets of SU(3)$_c$ and either singlets, doublets or triplets of SU(2)$_L$. Moreover, as argued in the main text, $\Psi$ need to be in odd SU(2)$_L$ representations for neutrino masses to be
        successfully generated at the two-loop level as in Fig.~\ref{fig:neutrino} of the main text. Thus, there are only two cases to consider regarding SU(2)$_L$, isosinglets or isotriplets.
        Note that since the fermions $\Psi_{L,R}$ now carry hypercharge, so the scalars $\eta,\chi_i$ also need to carry appropriate hypercharges as we will discuss shortly. 
	
In order for the $\Psi$ to mix with quarks, their hypercharge needs to be either $Y=-1/3$ or $Y=2/3$ leading to exotic-ordinary quark mixing. However, such a mixing in general will lead to two-body proton decays mediated by the scalar leptoquarks present in the model. For example the $p \to \pi^0 + e^+$ decay can happen through $\chi_2$ mediation\footnote{Note that in order to generate neutrino masses, the hypercharges of $\chi_{1,2}$ must be non-zero in this case, see Table.~\ref{tab:modelYneq0}.}. The non-observation of this decay channel sets a very stringent bound on the proton decay lifetime $\tau_p > 1.6 \times 10^{34}$ years, which requires the exotic scalar mediator mass $m_{\chi_2} \gsim 10^{15}$ GeV, to be very heavy near the typical GUT scale~\cite{Super-Kamiokande:2016exg}.
Note that, the typical colored scalar masses are given by $m_{\zeta}=\sqrt{\lambda_{\text{eff}}} f_{\text{PQ}}$.Therefore, the proton decay bounds sets a heavy mediator mass scale at odds with the post-inflationary DM scenario which requires~$f_a \lesssim 2 \times 10^{11}$ GeV. 
	
\begin{figure}[t!]
		\centering
		\includegraphics[scale=0.58]{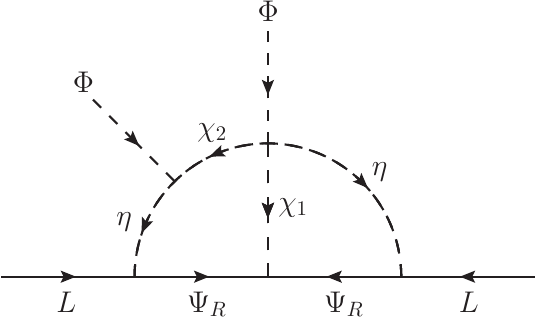} \hspace{+0.1cm}
		\includegraphics[scale=0.58]{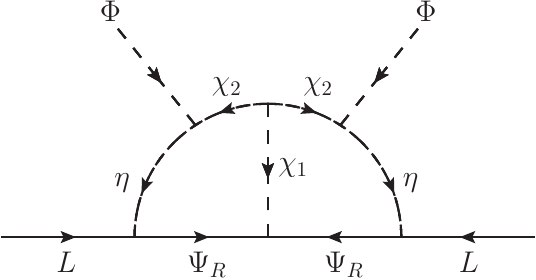} \hspace{+0.1cm}
		\includegraphics[scale=0.58]{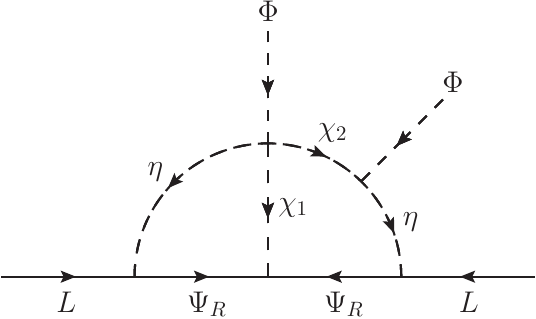}
		\caption{ Two-loop diagrams for neutrino mass generation mediated by the colored particles of Table~\ref{tab:modelYneq0}.}
		\label{fig:neutrinoYneq0}
	\end{figure}

The problem of proton decay in the hypercharge non-zero models can be avoided, by modifying the charges of the scalars $\eta, \chi_i$ under the PQ symmetry, so that after the breaking of the PQ symmetry, an appropriate residual symmetry survives unbroken, forbidding the proton decay. We now illustrate this idea by an example. The ${\rm SU(3)}_c \otimes {\rm SU(2)}_L \otimes {\rm U(1)}_Y \otimes {\rm U(1)}_{\rm PQ}$ charges of the particles are given in Table.~\ref{tab:modelYneq0}.   
	\begin{table}[t!]
		\renewcommand*{\arraystretch}{1.6}
		\centering
		\begin{tabular}{| K{2.5cm} || K{1.5cm} | K{5cm} |  K{1.5cm} |K{2.5cm} |}
			\hline
			&Fields&\SM&    U($1$)$_{\text{PQ}}$ & Multiplicity\\
			\hline \hline
			\multirow{2}{*}{Fermions}
			&$\Psi_L$&($\mathbf{3},\mathbf{1}, -1/3 \ (2/3)$)& $0$  & $n_\Psi$\\
			&$\Psi_R$&($\mathbf{3},\mathbf{1}, -1/3 \ (2/3)$)&  $-1/n_\Psi$ & $n_\Psi$ \\
			\hline \hline
			\multirow{3}{*}{Scalars}
			&$\sigma$&($\mathbf{1},\mathbf{1}, 0$)& $1/n_\Psi$ & $1$\\
			&$\eta$&($\mathbf{3},\mathbf{2}, 1/6 \ (7/6)$)& $-1/n_\Psi$ & $n_\eta $\\
			&$\chi_1$&($\mathbf{3},\mathbf{1}, 2/3 \ (-4/3)$)& $2/n_\Psi$ & $n_{\chi_1}$ \\
			&$\chi_2$&($\mathbf{3},\mathbf{1}, -1/3 \ (2/3)$)& $-1/n_\Psi$ & $n_{\chi_2}$ \\
			\hline
		\end{tabular}
		\caption{ Matter content and charges for the KSVZ model with two-loop neutrino masses featuring exotic fermion mixing with SM down (up) quarks.}
		\label{tab:modelYneq0} 
	\end{table}
	Note that in Table.~\ref{tab:modelYneq0} the U$(1)_Y$ charges outside parentheses will lead to mixing between down-type quarks and $\Psi$,
        while those inside parentheses will lead to mixing between up-type quarks and $\Psi$.
        With these modified charges, the neutrino mass can be generated in a way anologous to the the one used in main text via the two-loop diagrams given in Fig.~\ref{fig:neutrinoYneq0}.
	The relevant new Yukawa terms are given by:  
	\begin{align}
		- \mathcal{L}_{\text{Yuk.}} \supset \mathbf{Y}_{\Psi} \overline{\Psi_L} \Psi_R \sigma + \frac{1}{2} \mathbf{Y}_{\chi_{1 j}} {\Psi^T_R}~C~\chi_{1 j} \Psi_R + \mathbf{Y}_i \overline{L} \eta^\ast_i \Psi_R + \mathbf{m} \overline{\Psi_L} d_R (+ \mathbf{m} \overline{\Psi_L} u_R) + \text{H.c.} ,
		\label{eq:LYukgenYneq0}
	\end{align}
	where $d_R$ ($u_R$) are SM down (up) quark singlets. The scalar-potential terms triggering neutrino mass generation are,
	\begin{align}
		V \supset \mu_{ijk} \chi_{1 i} \chi_{2 j}\chi_{2 k} + \kappa_{ij} \eta_i^\dagger \Phi  \chi_{2 j} + \lambda_{ijk} \Phi^\dagger \eta_i \chi_{1 j} \chi_{2 k} + \text{H.c.} \; .
		\label{eq:VneutrinoYneq0}
	\end{align}
	A few comments on these models are in order:
	\begin{itemize}
		\item Three distinct colored scalar species $\eta, \chi_1, \chi_2$ are required to successfully generate two-loop Majorana neutrino masses, which is one more than in the framework presented in the Letter, when $\Psi$ has $Y=0$.  
		
		\item The minimal multiplicities required for non-zero neutrino masses is $n_{\Psi}=2 \ (1), \ n_\eta= 1 \ (2), \ n_{\chi_1} = n_{\chi_2} = 1$, 
                  where only the left and right diagrams in Fig. \ref{fig:neutrinoYneq0} contribute to neutrino masses and there will be a massless neutrino.
                  All diagrams in Fig. \ref{fig:neutrinoYneq0} contribute to neutrino mass generation for $n_{\chi_2}>1$ and having $n_\eta>2$ or $n_\Psi>2$ leads to three massive light neutrinos.  
		
		\item For $Y=-1/3$ ($Y=2/3$) the term $\overline{\Psi_L} d_R$ ($\overline{\Psi_L}  u_R$) is present in the Lagrangian. This allows for $\Psi$ to decay into ordinary matter. Thus, if $m_\Psi < m_{\eta,\chi_1,\chi_2}$, the lightest colored particle mediating neutrino masses is no longer stable and one can envisage a post-inflationary axion DM scenario.
		
\item These models feature a domain wall number $N=1$. Therefore, the post-inflationary axion DM production will be dominated by the misalignment mechanism. As argued in the main text, this leads to a prediction for the initial misalignment angle of $\left<\theta_0^2\right> \simeq 2.15^2$. Hence, if $\Omega_a h^2 = \Omega_{\text{CDM}} h^2$, to ensure that DM is not over produced, $f_a \lesssim 2 \times 10^{11}$ GeV, which is at the reach of haloscope experiments [see Fig.~\ref{fig:gagg} in the main text].
		
		\item The model with $Y=-1/3$ ($Y=2/3$), leads to $E/N=2/3$ ($E/N=8/3$), the same prediction for $g_{a\gamma\gamma}$ as DFSZ II (DFSZ I) [see Fig.~\ref{fig:gagg} in the main text].  
		
		\item Compared to the framework outlined in the Letter, these models require colored scalars that carry non-zero PQ charges, such that phenomenologically dangerous
                  terms that enable proton decay are absent from the Lagrangian.  
                  In fact, after symmetry breaking, an accidental unbroken $\mathcal{Z}_6$ symmetry remains in the Lagrangian where the fields transform as:
                  $(L,e_R) \to z^3 (L,e_R)$, $(Q, d_R, u_R, \Psi_{L,R}, \chi_1) \to z^2 (Q, d_R, u_R, \Psi_{L,R}, \chi_1)$ and $(\eta,\chi_2) \rightarrow z^5 (\eta,\chi_2)$,
                  with $z$ being the sixth root of unity. Note that, $Q$ denotes the SM quark doublets and $e_R$ the right-handed lepton singlets.
                  Hence the lowest dimension proton decay operators, $d u u e$, $Q Q u e$, $Q L d u$, $Q Q Q L$, are forbidden by symmetry. 
                  If no PQ charge is assigned to the colored scalars, for the exotic fermion mixing with SM down-type quarks case, the terms $d_R^c \chi_2 u_R$ and $u_R^c \chi_2^\ast e_R$ will be present
                  in the Lagrangian, leading to proton decay at tree-level, e.g. $p \to \pi^0 + e^+$. 
		
		\item The isotriplet version of these models only allows $\Psi_R$ to mix with SM quarks. For these cases there is no way to forbid dangerous proton decay interactions while simultaneously allowing exotic-ordinary quark mixing.
		
	\end{itemize}



\end{document}